\documentclass{iau}
\usepackage{graphicx}
\usepackage{natbib}

\title[Solar radionuclides] %% give here short title %%
{The birth environment of the solar system constrained by the relative abundances of the solar radionuclides}

\author[E. D. Young]   %% give here short author list %%
{Edward D. Young$^1$}

\affiliation
{$^1$Department of Earth, Planetary, and Space Sciences, \\University of California, Los Angeles
 \\ email: {\tt eyoung@epss.ucla.edu} \\[\affilskip]
}

\pubyear{2019}
\volume{345}
\jname{Origins: From the ProtoSun to the First Steps of Life, Volume 1}
\editors{Bruce G. Elmegreen, L. Viktor T\'oth, Manuel G\"udel, eds}
\begin{document}

\maketitle

\begin{abstract}
The relative abundances of the radionuclides in the solar system at the time of
its birth are crucial arbiters for competing hypotheses regarding the birth
environment of the Sun.  The presence of short-lived radionuclides, as
evidenced by their decay products in meteorites, has been used to suggest that
particular, sometimes exotic, stellar sources were proximal to the Sun's birth
environment.  The recent confirmation of neutron star - neutron star (NS-NS)
mergers and associated  kilonovae as potentially dominant sources of r-process
nuclides can be tested in the case of the solar birth environment using the
relative abundances of the longer-lived nuclides.  Critical analysis of the 15
radionuclides and their stable partners for which abundances and production
ratios are well known suggests that the Sun formed in a typical massive
star-forming region (SFR).  The apparent overabundances of short-lived
radionuclides (e.g.\, $^{26} {\rm Al}$, $^{41}{\rm Ca}$, $^{36}{\rm Cl}$) in
the early solar system appears to be an artifact of a heretofore
under-appreciation for the important influences of enrichment by Wolf-Rayet
winds in SFRs.  The long-lived nuclides (e.g.\, $^{238}{\rm U}$, $^{244}{\rm
Pu}$, $^{247}{\rm Cr}$, $^{129}{\rm I}$) are consistent with an average time
interval between production events of $10^8$ years, seemingly too short to be
the products of NS-NS mergers alone.  The relative abundances of all of these
nuclides can be explained by their mean decay lifetimes and an average
residence time in the ISM of $\sim200$ Myr. This residence time evidenced by
the radionuclides is consistent with the average lifetime of dust in the ISM
and the timescale for converting molecular cloud mass to stars.

%% add here a maximum of 10 keywords, to be taken form the file <Keywords.txt>
\end{abstract}

\firstsection % if your document starts with a section,
              % remove some space above using this command.

\section{Introduction}

  A worthy goal for studying the origins of the solar system is to establish whether or not the Sun formed in a typical environment and under typical conditions as judged by comparisons with star-forming regions in the Milky Way today.  Advantages of making this link between the birth of our solar system and that of other planetary systems in general are twofold. Firstly, a reconstruction of the birth environment of the Sun is relevant to assessing the probability that planetary systems analogous to our solar system exist, have existed, or will exist in the Milky Way Galaxy.  Secondly, if we establish the basic characteristics of the solar birth environment, we can test our understanding of star and planet formation in analogous environments using the solar system as the most well understood example.

 Among the many physical and chemical characteristics of the solar system, the relative abundances of the radionuclides may be some of the best clues to the solar birth environment. The initial relative abundances of many radionuclides in the solar system have been obtained through painstaking measurements of the isotopes themselves in meteorites, or, in the case of the short-lived radionuclides (SLRs), their decay products in meteorites.   Here we present two different paradigms for interpreting the meaning of the relative abundances of the radionuclides and their implications for reconstructing the environment in which our solar system formed.  The two views are in effect basis vectors for all models seeking to explain the solar abundances of the radionuclides.  The argument is made that initial abundances of SLRs are likely to be typical of massive star-forming regions (SFRs) similar to the Cygnus or Carina regions in our Galaxy today.

\section{Overview}

Theories for the provenance of the solar-system radionuclides can be divided into two broad categories.  In the first category, which we will refer to here as ``punctuated delivery'', the radiochemistry of the solar system is the result of the mingling of materials from a variety of discrete nucleosynthesis sources that are sometimes attributed to individual stars of specified masses and ages, including AGB stars, supernovae (SNe), and Wolf-Rayet stars \citep[e.g.,][]{Wasse96,Wasse2006,Lugar2014,Dwarkadas2017}.  The punctuated delivery scenarios emphasize the ``granularity'' \citep{Wasse96} of the interstellar medium in that a chance encounter with a single stellar source can dominate the inventory of a particular nuclide comprising pre-stellar material.  In this type of interpretation, the solar abundances of radionuclides are the result of a sequence of random discrete events (e.g., the explosion of a proximal supernova).

The second category of explanations for the solar abundances of the radionuclides appeals to widespread, quasi-continuous self-enrichment of massive SFRs \citep{Jura2013,Young2014,Fujimoto2018}.  In these scenarios, abundances of SLRs represent a steady-state  resulting from quasi-continuous production and losses in the SFRs over time.  The chemistry of the solar system therefore reflects the global evolution of the SFR in which the Sun formed.  Some models are of course a mix of the two categories \citep[e.g.,][]{Goune2012}, representing vectors in the two-dimensional cartesian space spanned by the two theories described here.   In what follows the essence of the two distinct types of scenarios are compared using  simple mathematical formulations.  The results are displayed in plots of the relative abundances of radionuclides normalized for differences in chemistry and differences in stellar production rates against their average lifetimes imposed by radioactive decay.  An eventual assessment of which of these two distinct types of scenarios is most likely will place constraints on the events leading up to the formation of the solar system, and thus the solar birth environment.

\section{Punctuated Delivery}

The punctuated nature, or ``granularity'' \citep{Wasse96}, of stellar nucleosynthesis events that could have seeded parental solar system material with nuclides can be described by an equation based on a geometric series summing individual nucleosynthesis events with an average temporal spacing $\delta t$ followed by a final actual free decay time $\Delta t$  \citep{Wasse2006,Lugar2014}

\begin{equation}
\frac{{N_{\rm{R}} }}
{{N_{\rm{S}} }} = \left[ {\frac{{P_{\rm{R}} }}{{P_{\rm{S}} }}\;\frac{\delta t }{T}\left( {1 + \frac{{\rm{e} ^{ - \delta t /\tau } }}{{1 - \rm{e} ^{ - \delta t /\tau } }}} \right)} \right]\exp ( - \Delta t/\tau )
\label{eq1}
\end{equation}

\bigskip
\noindent\rm where $N_i$ and $P_i$ are the number and production rates for radionuclides (R) and stable isotope partner (S), respectively, $T$ is the age of the Galaxy (taken to be 7.4 Gyr at the time of solar system formation), $\Delta t$ is in this case the time interval between the last event (LE) and the birth of the solar system, and $\tau$ is the mean life of R against radioactive decay. Dividing $N_{\rm R}$ by $N_{\rm S}$  eliminates the effects of element-specific chemistry when evaluating Equation \ref{eq1} for different radionuclides.  $N_{\rm R}/N_{\rm S}$ ratios are well known for the early solar system from precise measurements of the radioactive decay products of the SLRs and the relative abundances of the longer-lived nuclides in meteoritical materials \citep{Huss2009}.  Comparisons of different radionuclides are further facilitated by dividing the abundance ratio on the left-hand side of Equation \ref{eq1} by the production ratio on the right-hand side of Equation \ref{eq1}.  Production ratios are obtained from models of stellar nucleosynthesis \citep[e.g.,][]{Rausc2002,Woosl2007,Sukhbold2018}. The resulting quotient, $ \alpha(R) = (N_{\rm R}/N_{\rm S})/(P_{\rm R}/P_{\rm S})$, for different radionuclides depends on the radioactive mean lifetime $\tau$ and the duration of radioactive decay between events, all else equal.   Greater time intervals of decay following nucleosynthesis produce steeper trajectories in plots of $\alpha(R)$ vs $\tau$.  If the radionuclides inherited by the solar system from its parent molecular cloud were all of similar average age, $ \alpha(R) $ should vary only with $\tau$, with the shorter-lived nuclides being much less abundant relative to their production rates than the longer-lived nuclides.

\begin{figure}
\centering
\includegraphics{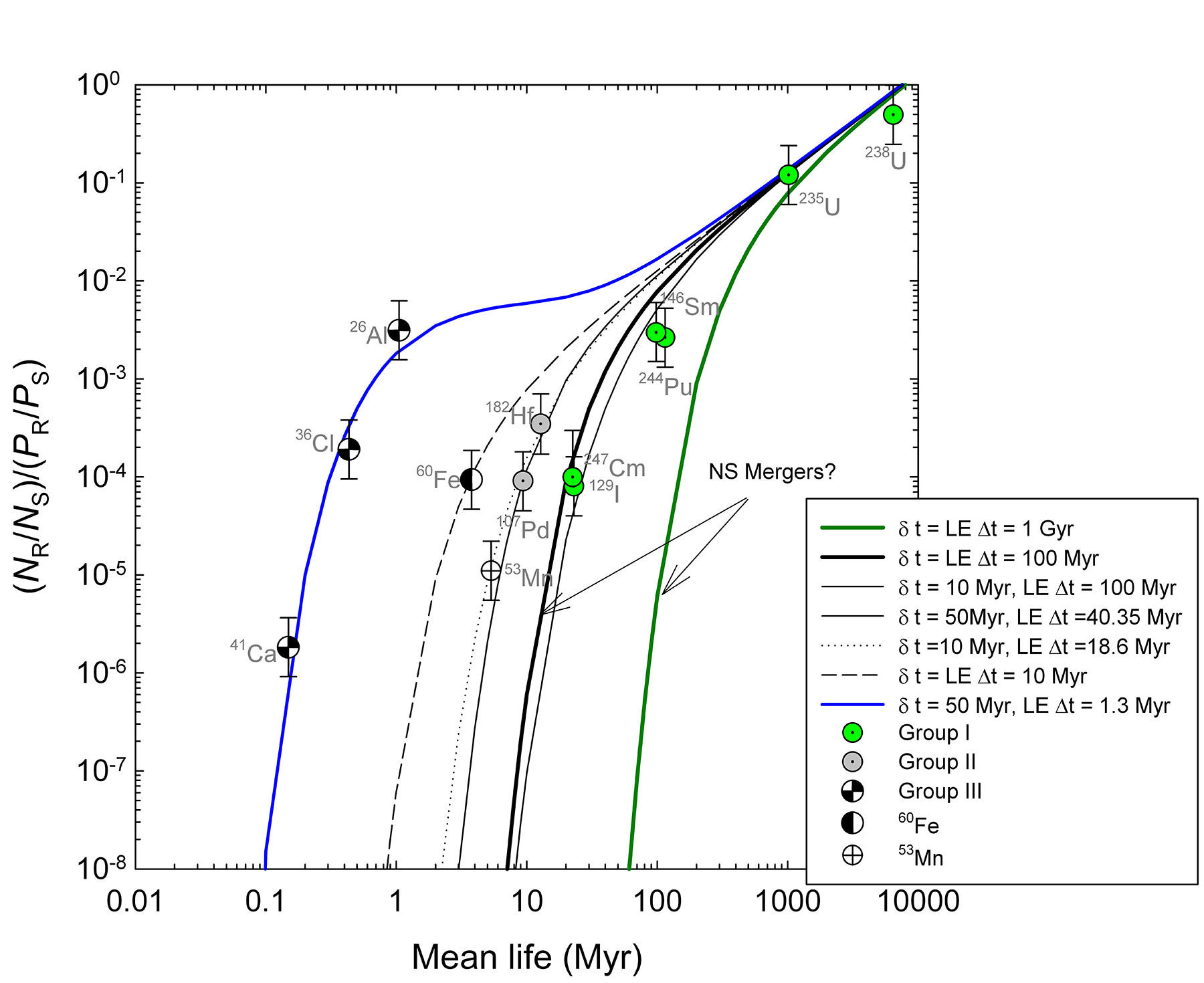}
\caption{Plot of radionuclide abundances ratioed to their stable nuclide partners ($N_{\rm{R}}/N_{\rm{S}}$) and production ratios ($P_{\rm{R}}/P_{\rm{S}}$) (or $\alpha(R)$ as described in the text) vs. mean decay lifetimes for radionuclides present at the birth of the solar system.  The values for the abundance ratios, production ratios, and decay lifetimes are from \cite{Young2014} and references therein updated with new production ratios for $\rm ^{182}Hf$ and $\rm ^{107}Pd$ \citep{Lugar2014} and the $\rm ^{247}Cm/^{232}Th$ ratio and production ratio from \cite{Tissot2016} and \cite{Goriely2001}, respectively. Model curves are from Equation \ref{eq1} fit to the different nuclide groups as indicated in the legend and in the text.}
\label{fig1}
\end{figure}

\cite{Young2016} showed that Equation \ref{eq1} can be used to group the solar radionuclides for which production ratios are well known.  Figure \ref{fig1} shows updated curves of $\log[(N_{\rm R}/N_{\rm S})/(P_{\rm R}/P_{\rm S})]$, or equivalently $\log({\alpha(R)})$, versus $\log({\tau})$, defining distinct groups of radionuclides.  These groupings succinctly summarize many of the relationships previously described in the literature \citep[e.g.,][]{Wasse2006,Lugar2014}.  For example, the relative concentrations of r-process neutron-capture products $\rm^{129}I$, $\rm ^{244}Pu$, $\rm ^{247}Cm$, $\rm ^{235}U$ and $\rm ^{238}U$ and the p-process nuclide $\rm ^{146}Sm$ are all explained by Equation (1) using $\delta t= 10$ Myr and $\Delta t = 100$ Myr (Figure \ref{fig1}, where $\rm ^{232}Th$ is used as the nearly stable nuclide to calculate $\alpha(R)$ for the actinides).  Here we label these isotopes as Group I. The value for $\delta t$ is consistent with the frequency of core-collapse supernova (CCSN) events affecting random positions in the Galactic disk \citep{Meyer2000} and so is appropriate if r-process nuclides form mainly in CCSNe.  Note that $\delta t$ is not the frequency of supernovae anywhere in the Galaxy, which is of order 1 per 50 yrs, but rather the frequency with which a random position in the Galaxy experiences a supernova explosion.

However, mounting evidence suggests that a principal source of r-process nuclides may be kilonovae resulting from mergers of neutron stars (NS-NS mergers) and possibly NS-black hole (NS-BH) mergers  \citep{Thielemann2017,Kasen2017,Smartt2017}.  Assuming similar production ratios for both CCSNe and NS-NS mergers \citep[but see][]{Cote2018}, especially among the actinides, Figure \ref{fig1} provides a constraint on the frequency of NS-NS mergers under the assumption that they were the primary source of solar r-process nuclides.  For example,  Figure \ref{fig1} shows two curves where the frequency of nucleosynthesis events is either 100 Myr or 1 Gyr (for simplicity in this case we set $\Delta t = \delta t$).  The latter $\delta t$ of 1 Gyr is in keeping with some estimates that NS-NS mergers are less frequent than CCSNe by a factor of 100 to 1000  \citep{Cote2017,Tsujimoto2014}.    The shorter event frequency $\delta t$ of 100 Myr fits the data reasonably well while the longer $\delta t$ of 1 Gyr does not fit the data.  The implication is that  either kilonovae events were only about ten times less frequent than CCSNe prior to solar system formation, or they were not the primary source of r-process nuclides leading up to the formation the Sun.

 \cite{Lugar2014} showed that a single set of values for $\delta t$  and $\Delta t$ can explain the relative concentrations of both s-process neutron addition nuclides  $\rm ^{107}Pd$ and $\rm ^{182}Hf$.  A value for $\delta t$ of 50 Myr, appropriate for AGB star encounters, and a similar value for $\Delta t$ of 40.35 Myr, fits the  $\log[(N_{\rm R}/N_{\rm S})/(P_{\rm R}/P_{\rm S})]$ values for $\rm ^{107}Pd$ and $\rm ^{182}Hf$ (Figure \ref{fig1}). We refer to these radionuclides as Group II.  $\rm ^{53}Mn$ and $\rm ^{60}Fe$ are treated separately from Group I and II. $\rm ^{53}Mn$  is also fit by the Group II curve but its origin must be distinct as it is a SN product, suggesting a shorter $\delta t$  interval.  $\rm ^{60}Fe$ requires its own $\Delta t$ (Figure \ref{fig1}). The shortest-lived nuclides, labeled Group III, have  $\log[(N_{\rm R}/N_{\rm S})/(P_{\rm R}/P_{\rm S})]$ values that are explained with Equation \ref{eq1} using $\delta t= 50$ Myr and $\Delta t = 1.3$ Myr (Figure 1).  The latter model reflects the fact that these short-lived nuclides have no "memory'' of events prior to their most recent synthesis.  Five distinct models defined by five $\Delta t$ values ($\delta t$ values are prescribed {\it a priori} by astrophysical constraints) represented by five curves are therefore required to explain the radionuclide abundances in Figure 1, one each for Groups I, II, and III and two others for $\rm ^{53}Mn$ and $\rm ^{60}Fe$.

 In general, quantitative assessments of punctuated additions of SLRs to regions of star formation lead to relatively low probabilities for the initial solar abundances of SLRs \citep{Adams2010, Adams2014}. In these scenarios, the solar system is assessed to be atypical in its complement of $\rm ^{26}Al$, for example.

\section{Quasi-continuous Self-enrichment of Star Forming Regions}

\begin{figure}
\centering
\includegraphics[scale=0.75]{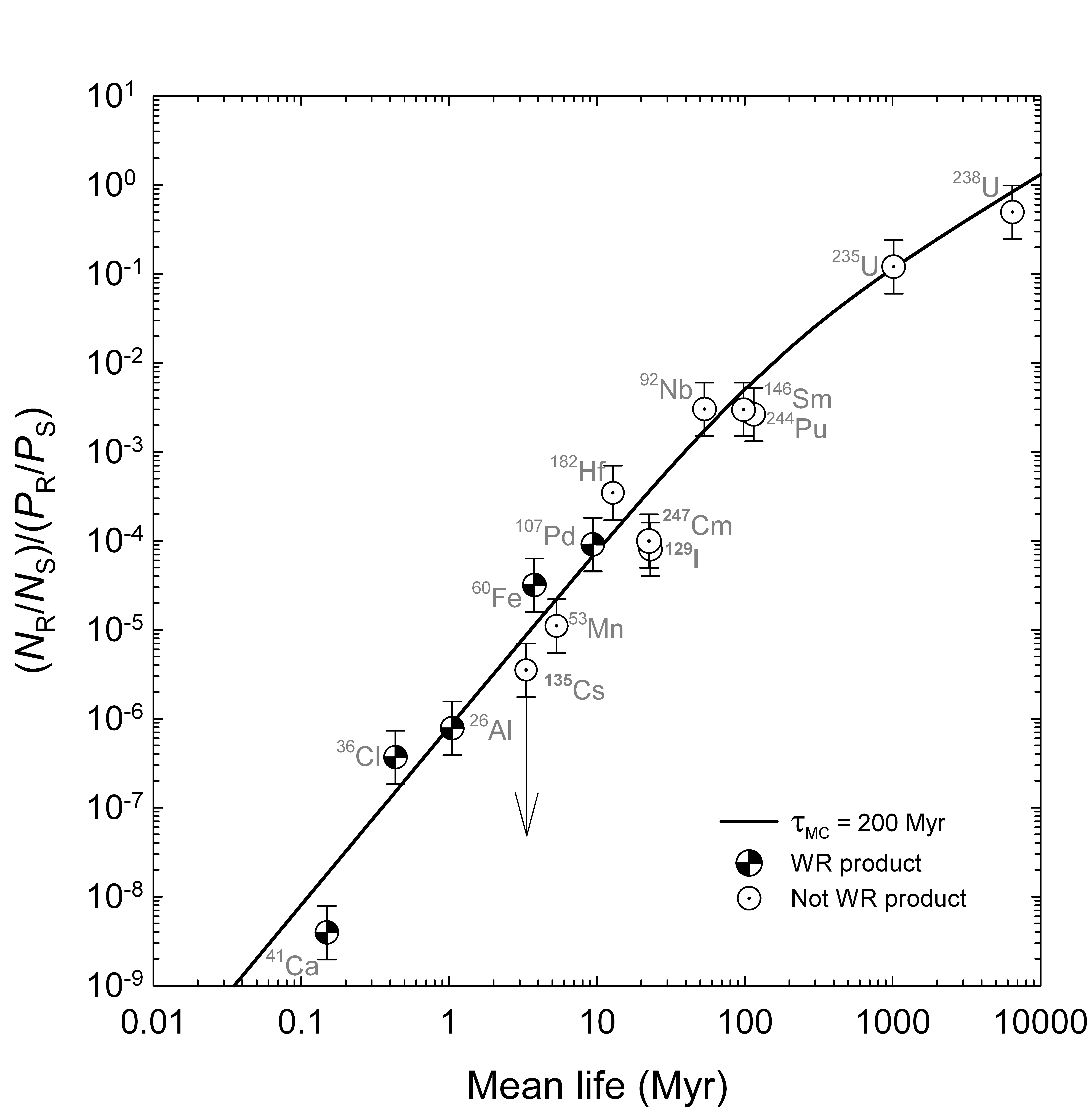}
\caption{Plot of radionuclide abundances normalized to their stable nuclide partners ($N_{\rm{R}}/N_{\rm{S}}$) and production ratios ($P_{\rm{R}}/P_{\rm{S}}$) vs. mean decay lifetime for radionuclides in the early solar system.  The plot is similar to Figure \ref{fig1} but with ordinate values calculated using production ratios that include $\Lambda_{\rm W}/\Lambda_{\rm SNe}$ from Equation \ref{eq3},  and published WR production terms compiled by \cite{Young2014}.  The curve is from Equation \ref{eq2} fit to the data using $\Lambda_{\rm W}/\Lambda_{\rm SNe} = 4000$ and $\tau_{\rm MC} = 200$ Myr with a reduced $\chi^2$ of $0.95$.  The value for $\rm ^{92}Nb$ and the upper limit for $\rm ^{135}Cs$ were not used to derive the best-fit curve, as discussed in the text.}
\label{fig2}
\end{figure}

The quasi-continuous self-enrichment model for the solar birth environment  relies on the enhanced production of several short-lived radionuclides emanating from Wolf-Rayet (WR) stars to explain the solar abundances of $\rm ^{26}Al$, $\rm ^{36}Cl$, and $\rm ^{41}Ca$ at the time the solar system formed (referred to as the initial abundances).  WR stars evolve from  massive progenitors ($M_{*} > \sim 25 M_{\odot}$ ).  Their large progenitor masses ensure that the WR stage occurs within several Myr of the birth of the star cluster.  WR stars are therefore expected to be spatially correlated with star-forming regions since they don't have time to flee their birth environments before they die \citep{Young2014}.  \cite{Young2014} showed that  the solar radionuclide abundances considered in Figure 1 are explained by a single model based on a two-phase interstellar medium (ISM) composed of a molecular cloud phase (MC) and a diffuse phase and with a molecular cloud mass fraction $x_{\rm MC}$ of  $\sim 0.17$, equivalent to that today.  The model equation  \citep{Young2014,Jacob2005} is

\begin{equation}
  \log \left( {\frac{{{N_{{\rm{R,MC}}}}}}{{{N_{{\rm{S,MC}}}}}}} \right) - \log \left( {\frac{{{P_{\rm{R}}}}}{{{P_{\rm{S}}}}}} \right) = 2\log \tau-
  \log \left[ {(1 - {x_{{\rm{MC}}}}){\tau _{{\rm{MC}}}} + \tau} \right] - \log {T} \\
 \label{eq2}
\end{equation}

\noindent  where the abundances of radionuclides and their stable partners now refer to those in the molecular cloud phase and environs in the SFR  ($N_{\rm{R,MC}}$ and $N_{\rm{S,MC}}$, respectively) as opposed to the diffuse ISM outside of the SFR.  The radionuclide production terms relevant to the molecular cloud setting are cast in terms of production from supernovae ( $P^{\rm{SNe}}_{\rm{R}}$) and production from WR winds ($P^{\rm{W}}_{\rm{R}}$):

\begin{equation}
\frac{{{P_{\rm{R}}}}}{{{P_{\rm{S}}}}} = \frac{{{\Lambda _{{\rm{SNe}}}}P_{\rm{R}}^{{\rm{SNe}}}}}{{{\Lambda _{{\rm{SNe}}}}{P_{\rm{S}}}}} + \frac{{{\Lambda _{\rm{W}}}P_{\rm{R}}^{\rm{W}}}}{{{\Lambda _{{\rm{SNe}}}}{P_{\rm{S}}}}}
\label{eq3}
\end{equation}

\bigskip
\noindent where $\Lambda_{\rm W}$ and $\Lambda_{\rm SNe}$ are the relative efficiencies for trapping the two sources of nuclides in the star-forming region.  WR production values are listed by  \cite{Young2014}.

\cite{Young2016} showed that the model fits solar-system data using two independent parameters, an enhancement of WR wind production over SN production in SFRs, with $\Lambda_{\rm W}/\Lambda_{\rm SNe} \sim 4000$ , and a sequestration time of nuclides in molecular cloud dust,  $\tau_{\rm MC}$, of $\sim 200$ Myr (Figure \ref{fig2}).  The $\tau_{\rm MC}$ in Equation \ref{eq2} is the {\it average} time spent in SFR molecular clouds where they experience radioactive decay with no new additions before incorporation into stars. Because individual clouds exist for shorter time spans than the SFR as a whole, time spent passing from one cloud to another via inter-cloud space in the SFR is included in the residence time. The radionuclide residence time $\tau_{\rm MC}$ of $\sim 200$ Myr  is consistent with the lifetime of dust in the interstellar medium \citep{Tiele2005} and simple estimates for the timescale for converting molecular clouds to stars \citep{Drain2011}.  In general, large values for $\Lambda_{\rm W}/\Lambda_{\rm SNe}$ are consistent with the fact that massive stars like WR progenitors apparently do not typically end their lives as energetic SNe but rather collapse by fallback to form black holes directly \citep{Fryer99,Smart2009}.  In part for this reason, the products of WR winds are enhanced relative to SNe products in SFRs sufficiently massive as to host a population of large stars \citep{Young2014}. The relatively quiescent ultimate fate of WR stars is supported by calculations showing that the fraction of stars with $M_{*} > 18 M_{\odot}$ that explode rather than undergoing direct collapse is just $~ 8\%$ \citep{Sukhbold2018}.

Numerical simulations \citep[e.g., ][]{Young2014} show that if clusters form every 1 to $1.5$ Myr in an SFR, and if mixing is sufficiently efficient, then the abundances of $\rm ^{26}Al$, $\rm ^{36}Cl$, and $\rm ^{41}Ca$ will be enhanced in SFRs relative to the average ISM as a whole.  \cite{Jura2013} appealed to this concept when they pointed out that the initial solar relative abundance of $\rm ^{26}Al$, usually expressed as $\rm ^{26}Al/\rm ^{27}Al = 5\times 10^{-5}$, is indistinguishable from that deduced by assuming that the 1.5 to 2.5 $M_{\odot}$ of $\rm ^{26}Al$ in the Galaxy \citep{Marti2009} resides primarily in regions where stars are forming.  These SFRs are  represented to first order by the $8.4\times 10^8 M_{\odot}$  of $\rm H_2$ gas in the Galaxy \citep{Drain2011} and solar-like ratios of heavy elements. A correlation between SFRs and $\rm ^{26}Al$ in the Galaxy is supported by maps of the $1.8$ MeV gamma ray emission from $\rm ^{26}Al$ decay in the Galaxy \citep{Diehl2006}.  The conclusion from this analysis is that the solar initial abundance of $\rm ^{26}Al$, and by inference those of $\rm ^{36}Cl$ and $\rm ^{41}Ca$ as well, was not extraordinary, but rather typical of star-forming regions as the result of self enrichment by WR stars. A recent numerical simulation at the Galactic scale by \cite{Fujimoto2018} leads to a similar conclusion regarding the importance of widespread self enrichment of SLRs in and near giant molecular clouds in the Galaxy.

The earlier work showing that a single curve in Figure \ref{fig2} explains the solar abundances of radionuclides makes the tacit prediction that as more radionuclide initial abundances are obtained from measurements in meteorites, they too should fall on the same curve defined by the two parameters $\Lambda_{\rm W}/\Lambda_{\rm SNe} \sim 4000$ and $\tau_{\rm MC} \sim 200$ Myr.  This is because these two parameters should be intrinsic properties of the SFR in which the solar system formed.  This prediction is put to the test by new data for the initial relative abundances of two additional short-lived radionuclides.  These new data include an estimate for the initial abundance of the p-process product $\rm ^{92}Nb$, yielding $\rm ^{92} Nb/^{93}Nb = 1.7\pm 0.6 \times 10^{-5}$  \citep{Iizuka2016}, and a new upper limit on the initial abundance of the s-process product $\rm ^{135}Cs$, yielding $\rm ^{135}Cs/^{133}Cs \le 2.8\times 10^{-6}$  \citep{Brennecka2017}.  Using the mean production ratio for $\rm ^{92}Nb$/$\rm ^{93}N$ of $5.65 \times 10^{-3}$ from \cite{Schonbachler2002} and the $\rm ^{135}Cs/^{133}Cs$ production ratio of 0.8 from \cite{Harper1996}, the new data plot within error of the solar-system curve in Figure \ref{fig2}.  The new data are thus supportive of the prediction that all radionuclides, excluding those produced by spallation, will plot on the curve defined by $\Lambda_{\rm W}/\Lambda_{\rm SNe} \sim 4000$ and $\tau_{\rm MC} \sim 200$ Myr.  If this proves to be true as more data are acquired, we will have learned about two fundamental parameters that characterized the SFR in which the solar system formed.

\section{Model Selection and Concluding Remarks}

The two different interpretations of the solar-system radionuclide data portrayed in Figures \ref{fig1} and \ref{fig2} each have elements to recommend them. For example, the physical and temporal separation between  r-process nucleosynthesis and production of lower-mass nuclides  by CCSNe  implied by the new kilonova data might be expressed by the different curves for these nuclides in Figure \ref{fig1}.  Conversely, all 15 radionuclides for which we have good estimates of $\alpha(R)$ values can be fit with a single curve with 2 fit parameters in Figure \ref{fig2} with an acceptable reduced $\chi^2$ of near unity.  Parsimony would seem to be on the side of the quasi-continuous self-enrichment model.  This point has been made using a simple Bayesian analysis \citep{Young2016}.  However, parsimony may not be the best arbiter for choosing between the two models described here.  Nonetheless, if the self-enrichment model continues to explain more data coming from detailed analyses of meteorite samples, the physical and chemical characteristics of the solar system may prove useful for understanding better the nature of massive star-forming regions in general.

\begin{discussion}

\discuss{Lissauer} {Assuming Wolfe-Rayet stars are the primary source of $^{26}$Al
in the Solar System, what fraction of planetary systems would you expect to have
similar excesses?}

\discuss{Young} {A consequence of discrete sampling of the stellar initial mass
function is that larger star clusters more reliably produce more massive stars.
Larger clusters implies formation in higher-mass star-forming regions.  The
fraction of planetary systems with solar-like relative concentrations of $\rm
^{26}Al$ should therefore correlate with the fraction of systems that form in
massive star-forming regions where the high-mass end of the IMF is well sampled
and WR stars are virtually guaranteed. }

\discuss{Kruijssen} {What is the nature of the molecular cloud ``residency
timescale'' that you use to explain the radioisotope abundance as a function of
their lifetime? You seem to use something like the gas depletion time (or
Mgas/SFR), but cloud lifetimes are now being found to be 10-30 Myr in our recent
work. How would the curve change if you use a shorter timescale?}

\discuss{Young} { We consider that characteristic evolutionary timescales correlate with scale, such that of order $10^6$ yrs  applies to formation of individual clusters, $\sim 3\times 10^7$ yrs applies for the longevity of giant molecular clouds, and $10^{8}$ yrs applies to the typical lifetime of molecular cloud complexes comprising spiral arms.  Of these timescales, the gas depletion time is, arguably, the best measure of the time that a typical nuclide persists from its synthesis to its incorporation into a star and associated planetary system. The relative abundances of the solar-system radionuclides, especially the long-lived nuclides, are telling us that the average interval for isolation and radioactive decay was on the order of 200 Myr leading up to the formation of the solar system.  This implies that these nuclides were likely passed from cloud to cloud as a result of inefficient star formation in the clouds.  Evolution curves in Figure \ref{fig2} would be considerably shallower with radioactive free decay times of 10 to 30 Myr, and would not be consistent with the long-lived nuclide data irrespective of the implications for the short-lived nuclides. }

\end{discussion}

\end{document}